\documentstyle[12pt,epsfig]{article}
\def \medio  {\baselineskip= 1.5 \normalbaselineskip}

\newcommand{\titul}[1] {\begin{center}{\large\bf #1 } \end{center}\vskip 1.cm}
\newcommand{\autor}[1] {\begin {center} {\large \lineskip .5em #1 }
                        \end   {center} }
\newcommand{\lugar}[1] {\begin{center} {\it #1} \end{center}}
\newcommand{\abstr}[1] {{\begin{center} \vskip .5cm {\bf Abstract
                        \vspace{0pt}} \end{center}}\begin{quote} #1
                        \end{quote}}
\newcommand{\z}{&&\hspace*{-1cm}}

\newcommand{\bea}{\begin{eqnarray}}
\newcommand{\eea}{\end{eqnarray}}
\begin{document}
\begin{titlepage}

\begin{flushright} {
%8th version \\
%\bf US-FT/11-99 \\ 
%March 23,
July, 2002 \\
hep-ph/0207226
} \end{flushright}

%\vskip 3.cm
\titul{ The 
 contribution of off-shell gluons\\
 to the longitudinal structure function $F_L$\\
% and the unintegrated gluon distributions
}

\autor{
A.V. Kotikov
%\footnote{E-mail:kotikov@sunse.jinr.ru}
}
\lugar{Bogoliubov Laboratory of Theoretical Physics\\
Joint Institute for Nuclear Research\\
141980 Dubna, Russia}
\autor{A.V. Lipatov}
\lugar{Department of Physics \\
%M.V. 
Lomonosov Moscow State University\\
119899 Moscow, Russia}
%\autor{G. Parente}
%%\footnote{E-mail:gonzalo@fpaxp1.usc.es}
%\lugar{Departamento de F\'\i sica de Part\'\i culas\\
%Universidade de Santiago de Compostela\\
%15706 Santiago de Compostela, Spain}
\autor{N.P. Zotov}
\lugar{
%D.V. 
Skobeltsyn Institute of Nuclear Physics \\
%M.V. 
Lomonosov Moscow State University\\
119992 Moscow, Russia}
\abstr{
\medio
We present results for
the structure function $F_L$ 
for a gluon target having nonzero transverse
momentum square at order $\alpha _s$.
The results of double convolution 
(with respect to Bjorken variable
$x$ and the
transverse momentum) of the perturbative part and the unintegrated
gluon densities are compared with recent experimental data for
$F_L$ at low $x$ values and with predictions of other approaches.

\vskip 0.5cm

PACS number(s): 13.60.Hb, 12.38.Bx, 13.15.Dk

}
\end{titlepage}
\newpage

\pagestyle{plain}
%\medio
\section{Introduction} \indent 

The basic information on the internal
%quark
structure of nucleons is extracted from the process of deep inelastic
(lepton-hadron) scattering (DIS). Its differential cross-section has the form:
\begin{eqnarray}
 \frac{d^2 \sigma}{dxdy}~=~ \frac{2 \pi \alpha_{em}^2}{xQ^4}~~ \bigl[
\left( 1 - y + y^2/2
\right) F_2(x,Q^2) - \left(y^2/2\right) F_L(x,Q^2) \bigr],
%\label{0.1} 
\nonumber \end{eqnarray}
where $F_2(x,Q^2)$ and $F_L(x,Q^2)$ are the transverse and
longitudinal structure functions (SF), respectively,
$q^{\mu}$ and  $p^{\mu}$ are the photon and the hadron 4-momentums and
$x=Q^2/(2pq)$ with $Q^2 = -q^2>0$. 

The longitudinal SF $F_L(x,Q^2)$ 
is a very sensitive QCD characteristic because it is
equal to zero in the parton model with spin$-1/2$ partons.
Unfortunately, essentially at
small values of $x$, the experimental extraction of $F_L$ data 
is required a
rather cumbersome procedure (see \cite{1.5,BaGlKl}, for example).
Moreover, the perturbative QCD leads to some controversial results in the 
case of SF $F_L$.
The next-to-leading order (NLO) corrections to the longitudinal 
coefficient function, which are 
large and negative at small $x$ 
\cite{Keller,Rsmallx}, need a resummation procedure 
\footnote{ Without a resummation the NLO
%LO$\&$NLO 
approximation of SF $F_L$ 
can be negative at 
low $x$ and quite low $Q^2$ values (see \cite{Rsmallx,Thorne02}).}
that leads to 
coupling constant scale higher essentially then $Q^2$ 
(see \cite{DoShi,Rsmallx,Wong})
\footnote{Note that at low $x$
a similar property has been observed
also in 
%BFKL-motivated 
the approaches  \cite{BFKLP,bfklp1,Salam}
(see recent review \cite{Andersson} and discussions therein),
which based on Balitsky-Fadin-Kuraev-Lipatov (BFKL) dynamics
\cite{BFKL}, where the leading $\ln(1/x)$ contributions are summed.}.

Recently there have been important new data \cite{H1FL97}-\cite{Gogi}
of the longitudinal SF
%structure function (SF) 
$F_L$, 
which have probed the small-$x$ region down to $x \sim 10^{-2}$. 
Moreover, the SF $F_L$  can be related at small $x$ with SF $F_2$ 
and the derivation $dF_2/d\ln(Q^2)$ (see \cite{KoFL}-\cite{KoPaR}). In 
this way most precise predictions based on data of $F_2$ 
and $dF_2/d\ln(Q^2)$ (see \cite{H1FL} and references therein)
 can be obtained for $F_L$.
% (see Section 4). 
These predictions can be 
considered as indirect 'experimental data' for $F_L$.\\

In this paper for analysis of the above data
%, where the $x$ values are quite small 
 we will use so called
$k_T$-factorization approach \cite{CaCiHa, CoEllis, LRSS}
based on BFKL dynamics \cite{BFKL}
(see also recent review \cite{Andersson} and references therein).
In the framework of the $k_T$-factorization approach, a study of the 
longitudinal SF $F_L$ has been done firstly in Ref. \cite{CaHa},
where small $x$ asymptotics of $F_L$ has been evaluated
using the BFKL results for  the Mellin transform of the
unintegrated gluon distribution and the longitudinal Wilson coefficient 
functions has 
been calculated analytically
for the full perturbative series at asymptotically small
$x$ values. Since we want to analyze 
%experimental data for the SF
$F_L$ data
%, we have an interest to obtain results at quite 
in a broader  range at
%range of 
small $x$,
% values. For the reason 
we will use parameterizations
of the unintegrated gluon distribution function $\Phi_g(x,k^2_{\bot})$
(see Section 3).

The  unintegrated gluon distribution
$\Phi_g(x,k^2_{\bot})$ 
%$\Phi(x_B,k^2_t)$ 
($f_g$ is the (integrated) gluon distribution in proton multiplied
by $x$ and $k_{\bot}$ is the transverse part of gluon 4-momentum $k^{\mu}$)
 \begin{eqnarray}
f_{g}(x,Q^2) ~=~ \int^{Q^2}dk^2_{\bot}
\, \Phi_g(x,k^2_{\bot}) 
~~~~~\mbox{(hereafter } 
%~q^2=-Q^2,
~k^2=-k^2_{\bot} \mbox{)}
\label{1}
 \end{eqnarray}
is the basic dynamical quantity in 
%BFKL 
the $k_T$-factorization approach \footnote{
In our previous analysis \cite{KLPZ} we shown
that the property
$k^2=-k^2_{\bot}$ leads to
%comes from 
the equality of the Bjorken $x$ value 
%of corresponding parton Bjorken variable $z=Q^2/(2kq)$ 
in the standard renormalization-group approach 
%(where $x=Q^2/2/(kq)$) 
and in the Sudakov one.}.
It
%and 
%which 
satisfies the BFKL equation \cite{BFKL}. 

Notice that the integral is divergent at  lower limit 
(at least, for some parameterizations of $\Phi_g(x,k^2_{\bot})$)
and so it leads to the
necessity to consider the difference $f_{g}(x,Q^2) - f_{g}(x,Q^2_0)$
with some nonzero $Q^2_0$ (see discussions in \cite{KLPZ}),
%Section 3), 
i.e.
 \begin{eqnarray}
f_{g}(x,Q^2) ~=~ f_{g}(x,Q^2_0) + \int^{Q^2}_{Q^2_0}
dk^2_{\bot} \, \Phi_g(x,k^2_{\bot}) 
\label{1.1}
 \end{eqnarray}

Then, in the $k_T$-factorization
the SF $F_{2,L}(x,Q^2)$ are driven at small $x$ primarily
by gluons and are related in the following way to the unintegrated 
distribution $\Phi_g(x,k^2_{\bot})$: 
\begin{eqnarray}
F_{2,L}(x,Q^2) ~=~\int^1_{x} \frac{dz}{z} \int 
dk^2_{\bot} \sum_{i=u,d,s,c} e^2_i
\cdot \hat C^g_{2,L}(x/z,Q^2,m_i^2,k^2_{\bot})~ \Phi_g(z, k^2_{\bot}), 
 \label{d1}
\end{eqnarray}
where $e^2_i$ are charge squares of active quarks.

The functions $
%\sum_{i=u,d,s,c} 
\hat C^g_{2,L}(x,Q^2,m_i^2,k^2_{\bot})$ 
can be regarded as 
%the structure functions 
SF of the 
off-shell gluons with virtuality $k^2_{\bot}$ (hereafter we call them as
{\it hard structure functions }
\footnote{by analogy with similar 
%corresponding 
relations  
%to (\ref{d1}) 
between cross-sections and hard 
cross-sections.}). They are described by the sum of the quark 
boxes (and crossed boxes) diagram contribution to the 
photon-gluon interaction 
(see Fig. 1). 

The purpose of the paper is to 
give predictions for the longitudinal SF $F_{L}(x,Q^2)$
based on the calculations of the hard SF
%coefficient functions
$\hat C^g_{2,L}(x,Q^2,m^2,k^2_{\bot})$, given in our previous study 
\cite{KLPZ},
and several parameterizations of unintegrated gluon distributions (see 
\cite{Andersson} and references therein).
%$F_{L}(x_B,Q^2)$.

It is instructive to note that
the diagrams shown in Fig. 1. are similar to those  of the
photon-photon scattering process.
The corresponding QED contributions have been calculated many years ago
in Ref. \cite{BFaKh} (see also the beautiful review in Ref. \cite{BGMS}). 
Our results 
have been calculated independently 
and they are in full agreement with Ref.
\cite{BFaKh}. 
Moreover, our results are in agreement with the corresponding integral 
representations for $\hat C^g_{2,L}$, given in \cite{CaCiHa,CaHa} and 
numerically with results of \cite{BMSS}. 
However, we hope
that our formulas which are given in a 
%more 
%have essentially 
simpler form 
%and we hope that they will 
could be useful for others.
This simpler form for the hard SF $\hat C^g_{2,L}$ comes from using 
the relation  between results based on non-sense, transverse 
and longitudinal gluon
%sets of 
polarizations (see Eq. (\ref{3dd}) below)
observed in \cite{KLPZ} for gauge-invariant sets of diagrams.\\

The structure of this paper is as follows: in Section 2 we
present the basic formulae of our approach.
Section 3 contains the relations between SF $F_L$ and $F_2$ and 
the derivative $dF_2/d\ln Q^2$, obtained in \cite{KoFL,KoPaFL,KoPaR} 
in the framework of 
Dokshitzer-Gribov-Lipatov-Altarelli-Parisi (DGLAP) approach \cite{DGLAP} 
(i.e. in the collinear approximation: 
$k^2_{\bot}=0$).
%perturbative QCD.
In Section 4 we give the predictions for the structure function
$F_L$ 
%and $F_L^c$ 
for three cases of unintegrated gluon distributions.

\section{Basic formulae} \indent

To begin with, we review shortly 
%In the Section we illustrate some 
results of Ref. \cite{KLPZ} needed below in
our investigations.

The hadron part of the DIS spin-average lepton-hadron 
cross section can be represented in the form 
\footnote{Hereafter we consider only one-photon exchange approximation.}
 \begin{eqnarray}
F_{\mu\nu} ~=~ e_{\mu\nu}(q)~F_{L}(x,Q^2) +  d_{\mu\nu}(q,p)
~F_{2}(x,Q^2),
 \label{2}
 \end{eqnarray}
where 
%$p^{\mu}$ are hadron momenta, $x_B=Q^2/(2pq)$,
 \begin{eqnarray}
e_{\mu\nu}(q) = g_{\mu\nu}- \frac{q_{\mu}q_{\nu}}{q^2}~~\mbox{ and }~~
d_{\mu\nu}(q,p) = -\Biggl[ g_{\mu\nu}+ 
2x_B\frac{(p_{\mu}q_{\nu}+p_{\nu}q_{\mu})}{q^2} + 
p_{\mu}p_{\nu}\frac{4x_B^2}{q^2} \Biggr]
\nonumber
\end{eqnarray}
%and
%$F_k(x,Q^2)~~($hereafter $k=2,L)$ are so-called structure functions.

\subsection{Feynman-gauge gluon polarization}
 \indent 
As it has been shown in Ref.
%our previous study 
\cite{KLPZ}, it is very
convenient to consider,
%{\bf 1.} 
as the first approximation,  
gluons having 
polarization tensor (hereafter
the indices $\alpha$ and $\beta$ are connected with gluons and 
$\mu$ and $\nu$ are connected with photons)\footnote{In 
principle, we can use here more general cases of polarization 
tensor (for example, one is based on Landau or unitary gauge). The difference
between them and (\ref{01dd}) is $\sim k^{\alpha}$ and/or $\sim k^{\beta}$ 
and, hence, it leads to zero contributions because Feynman diagrams on Fig.1
are gauge invariant.}: 
\bea
\hat P^{\alpha\beta}=-g^{\alpha\beta}
\label{01dd}
\eea
The 
%polarization 
tensor corresponds is equal to the standard choice of polarization matrix
in the framework of 
%the 
collinear approximation.
In a sense the case of polarization is equal to the standard DIS
suggestions about parton properties, excepting their off-shell property. 
The polarization (\ref{01dd})
%It 
gives the main
 contribution to the polarization tensor  we are interested in (see below)
\bea
\hat P^{\alpha\beta}_{BFKL}=
\frac{k_{\bot}^{\alpha}k_{\bot}^{\beta}}{k_{\bot}^2},
\label{1dd}
\eea
which
%It 
comes from the high energy (or $k_T$) factorization prescription
\cite{CaCiHa, CoEllis, LRSS}
\footnote{We would like to note that the BFKL-like polarization tensor 
(\ref{1dd}) is a particular case of so-called nonsense polarization of the 
%gauge 
particles in $t$-channel. The nonsense polarization
makes the main contributions for cross sections in $s$-channel at 
$s \to \infty$ (see, for example, Ref. \cite{KurLi} and references therein).
The limit $s \to \infty$ corresponds to the small values of Bjorken variable
$x$, that is just the range of our study.}.

Contracting the photon projectors (connected with photon indices of diagrams
on Fig.1.) 
 \begin{eqnarray}
\hat P^{(1)}_{\mu\nu} = -\frac{1}{2} g_{\mu\nu}~~\mbox{ and }~~
\hat P^{(2)}_{\mu\nu} = 4z^2\frac{k_{\mu}k_{\nu}}{Q^2}~,~~~~~
%\left(z=\frac{Q^2}{2(kq)}\right),
\nonumber
% \label{1}
 \end{eqnarray}
(here $z=Q^2/(2pq)$ is the corresponding Bjorken variable on parton level)
with the parton tensor $F^p_{\mu\nu}$:
 \begin{eqnarray}
F^p_{\mu\nu} ~=~ e_{\mu\nu}(q)~F^p_{L}(z,Q^2) +  d_{\mu\nu}(q,k)
~F^p_{2}(z,Q^2),
 \label{2p}
 \end{eqnarray}
%(here $k^{\mu}$ is the parton 4-momentum)
we obtain 
%the following relations from Eq.(\ref{1}) 
at the parton level (i.e. for off-shell gluons having 
momentum $k_{\mu}$), when $\hat C^g_{2,L}(z) \sim F^p_{2,L}(z,Q^2)$
\footnote{ The hard SF $\hat C^g_{2,L}$ do not depend on the type of target, 
so we can replace $z \to x$ below.}:
 \begin{eqnarray}
\tilde \beta^2 \cdot \hat C^g_{2}(x) &=& 
%e_c^2 \cdot \frac{\alpha_s(Q^2)}{4\pi}\cdot x 
{\cal K}
\cdot
\left[
f^{(1)} + 
\frac{3}{2\tilde \beta^2}\cdot f^{(2)} \right]
 \label{3}\\
\tilde \beta^2 \cdot \hat C^g_{L} (x) &=& 
%e_c^2 \cdot \frac{\alpha_s(Q^2)}{4\pi}\cdot x 
{\cal K}
\cdot
\left[
4bx^2 f^{(1)} + 
\frac{(1+2bx^2)}{\tilde \beta^2}\cdot f^{(2)} \right]
~=~
%e_c^2 \cdot \frac{\alpha_s(Q^2)}{4\pi}\cdot x 
{\cal K}
\cdot
f^{(2)} +4bx^2 \tilde \beta^2\cdot C^g_2 ,
 \label{4}
%\nonumber
 \end{eqnarray}
where the normalization factor 
${\cal K} =  %e_c^2 \cdot 
a_s(Q^2)
%\alpha_s(Q^2)/(4\pi)
\cdot x$, 
$$\hat P^{(i)}_{\mu\nu} F_{\mu\nu} =
%e_c^2 \cdot \frac{\alpha_s(Q^2)}{4\pi}\cdot x 
{\cal K} \cdot  f^{(i)} 
, \, i = 1, 2$$ and 
$a_s(Q^2)=\alpha_s(Q^2)/(4\pi),~~
\tilde \beta^2=1-4bx^2,~~b=-k^2/Q^2 \equiv  k_{\bot}^2/Q^2 >0,~~
a=m^2/Q^2$.

Applying the projectors $\hat P^{(i)}_{\mu\nu}$ to Feynman diagrams  
displayed in Fig.1, we obtain
the following results 
%for the contributions to expressions 
\begin{eqnarray}
f^{(1)} &=& -2 \beta \Biggl[ 1 - \biggl(1-2x(1+b-2a) \cdot [1-x(1+b+2a)] 
\biggr) \cdot f_1  
\nonumber \\
    &+& (2a-b)(1-2a)x^2 \cdot f_2  \Biggr],
 \label{5}\\
f^{(2)} &=& 8x\cdot \beta \Biggl[(1-(1+b)x)  
-2x \biggl(bx(1-(1+b)x)(1+b-2a) + a\tilde \beta^2 \biggr)\cdot f_1  
\nonumber \\
    &+& bx^2(1-(1+b)x) (2a-b) \cdot f_2  \Biggr],
\label{6} 
\end{eqnarray}
where
\bea
\beta^2=1-\frac{4ax}{(1-(1+b)x)}
\label{6.1d}
\eea
and 
%\footnote{ We use the variables as defined in Ref. \cite{Vog}.}
$$f_1=\frac{1}{\tilde \beta \beta} \cdot
\ln\frac{1+\beta \tilde \beta}{1-\beta \tilde \beta},
~~~~~f_2=-\frac{4}{1-\beta^2 \tilde \beta^2}$$ 
\vskip 0.5cm

\subsection{BFKL-like gluon polarization} \indent 

%{\bf 2.}
Now we take into account 
the BFKL-like gluon polarization (\ref{1dd}). As we 
shown in \cite{KLPZ}, the 
%BFKL-like 
projector $\hat P^{\alpha\beta}_{BFKL}$
can be represented as
\bea
\hat P^{\alpha\beta}_{BFKL} ~=~ -\frac{1}{2}
\frac{1}{\tilde \beta^4} \left[\tilde \beta^2 g^{\alpha \beta} 
-12 bx^2 \frac{q^{\alpha}q^{\beta}}{Q^2} \right]
\label{3dd} 
\eea

In the previous subsection we have already presented the contributions to
%coefficient functions 
hard SF using the first term in the brackets of the r.h.s. of
(\ref{3dd}). Repeating the above calculations with the projector
$\sim q^{\alpha}q^{\beta}$, we obtain the total contributions to
%coefficient functions 
hard SF which can be represented as the following shift of the  
results in Eqs.(\ref{3}) -
%,\ref{4},\ref{5}) and 
(\ref{6}):
\bea
\hat C^g_{2}(x)~ &\to& ~ \hat C^g_{2,BFKL}(x),~~~
\hat C^g_{L}(x)~ \to ~ \hat C^g_{L,BFKL}(x);
\nonumber \\
 f^{(1)}~ &\to& ~f^{(1)}_{BFKL}= 
\frac{1}{\tilde\beta^4}
\left[ \tilde \beta^2 f^{(1)}  ~-~
3bx^2 \tilde f^{(1)}\right] \nonumber \\
 f^{(2)} ~&\to&~ f^{(2)}_{BFKL}=
\frac{1}{\tilde\beta^4}
\left[ \tilde \beta^2 f^{(2)}  ~-~
3bx^2 \tilde f^{(2)}
\right],
\label{4dd}
\eea
where
\begin{eqnarray}
\tilde f^{(1)} &=& - \beta \Biggl[ \frac{1-x(1+b)}{x}  
-2 \biggl(x(1-x(1+b))(1+b-2a) +a \tilde \beta^2 \biggr) \cdot f_1  
\nonumber \\
    &-& x(1-x(1+b))(1-2a) \cdot f_2  \Biggr],
 \label{5dd}\\
\tilde f^{(2)} &=& 4 \cdot \beta ~(1-(1+b)x)^2 \Biggl[2   
- (1+2bx^2)\cdot f_1  
%\nonumber \\    &-& 
-bx^2 \cdot f_2  \Biggr],
\label{6dd} 
\end{eqnarray}

Notice that the general formulae are needed only to evaluate the charm
contribution to structure functions $F_2$ and $F_L$, i.e. $F_2^c$ and 
$F_L^c$. To evaluate
the corresponding light-quark contributions, i.e. $F_L^l$, 
we can use $m^2=0$ limit of above formulae.

\subsection{The case $m^2 =0$}

When $m^2 = 0$ the hard SF
%coefficient functions
 $\hat C^g_k(x)$ are defined by $f^{(1)}$, $f^{(2)}$,  $\tilde f^{(1)}$ and 
$\tilde f^{(2)}$ (as in
Eqs.(\ref{3}), (\ref{4}) and (\ref{4dd})) which can be represented as
 \begin{eqnarray}
f^{(1)} &=& -2 \Biggl[ 2 - 
\biggl(1-2x(1+b)+2x^2(1+b)^2 \biggr) \cdot  L(\tilde \beta)
\Biggr],
 \label{6.4} \\
& & \nonumber \\
f^{(2)} &=& 8x (1+b)(1-(1+b)x) \Biggl[1 -
2bx^2  \cdot  L(\tilde \beta)
\Biggr], \label{6.5} \\
\tilde f^{(1)} &=& - \frac{(1+b)(1-x(1+b))}{bx}
\Biggl[1 -
2bx^2  \cdot  L(\tilde \beta)
 \Biggr]
= - \frac{1}{8bx^2} f^{(2)},
 \label{6.4dd} \\
& & \nonumber \\
\tilde f^{(2)} &=& 4 (1-x(1+b))^2
\Biggl[ 3 - (1+2bx^2)  \cdot  L(\tilde \beta)
\Biggr]
\label{6.5dd}
\end{eqnarray}
and, thus, (see Eq. (\ref{4dd}))
 \begin{eqnarray}
\tilde \beta^4 \cdot f^{(1)}_{BFKL} &=& 
(-2)
\left(1-x(1+b)\right)
\Biggl[ 2 
\biggl(1-2x(1+b)+\frac{x^2(1-b)^2}{1-x(1+b)} \biggr)
\nonumber \\
&-& \biggl(1-x(1+b)-4x^3b(1+b) +\frac{x^2(1-b)^2}{1-x(1+b)} \biggr)
 \cdot  L(\tilde \beta)
\Biggr],
 \label{6.8}\\
& & \nonumber \\
\tilde \beta^4 \cdot f^{(2)}_{BFKL} &=& 8x (1-x(1+b))
\Biggl[ 1+b-18bx(1-x(1+b)) \nonumber \\
&+& 2bx 
\left(3-4x(1+b)+6bx^2(1-x(1+b)) \right)  \cdot  L(\tilde \beta)
\Biggr],
\label{6.5d3}
\end{eqnarray}
where
$$
L(\tilde \beta)=\frac{1}{\tilde \beta } \cdot
\ln\frac{1+\tilde \beta}{1- \tilde \beta} $$

%%%%%%%%%%%%%% 3  3  3
\section{Relations between $F_L$, $F_2$ and derivation of $F_2$
in the case of collinear approximation} \indent

%The 
Another information about the SF $F_L$ can be obtained in the 
%framework of 
collinear approximation (i.e. when $k_{\bot}^2 =0$) in the following way.

In the framework of perturbative QCD, there is the possibility to connecting
$F_L$ to $F_2$  and its derivation $dF_2/d\ln Q^2$ due the fact that at 
small $x$ the DIS structure functions 
depend only on two {\it independent} functions: the gluon distribution and
singlet quark one (the nonsinglet quark density is negligible at small $x$),
which in turn can be expressed in terms of measurable SF $F_2$ and its
derivation $dF_2/d\ln Q^2$.

In this way, by analogy with the case of the gluon distribution function (see
\cite{Prytz,KoPaG} and references therein), the behavior of
$F_L(x,Q^2)$ has been studied in \cite{KoFL}-\cite{KoPaR},
% atsmall values of $x$, 
using the 
%quite old 
HERA data \cite{F2H1,F2ZEUS} and the method \cite{method1}
\footnote{The method is based on previous investigations \cite{1.5,method}.}
of replacement of the Mellin convolution by ordinary products.
Thus, the small $x$ behavior of
the SF $F_L(x,Q^2)$ can be extracted directly from the measured values
of $F_2(x,Q^2)$ and its derivative without a cumbersome procedure (see
\cite{1.5,BaGlKl}).  These extracted values of $F_L$ may be well considered as
{\it new
small $x$ 'experimental data' of $F_L$}.
The relations can be violated by nonperturbative corrections like 
higher twist ones (see \cite{HT,KriKo}), which can be large exactly
in the case of SF $F_L$ \cite{CaHa,BaGoPe}. \\

Because $k_T$-factorization approach is one of popular nonperturbative
approaches used at small $x$, it is very useful to compare its predictions
with the results of \cite{KoFL}-\cite{KoPaR} based on
%lead to 
the relations
between SF $F_L(x,Q^2)$, $F_2(x,Q^2)$
and $dF_2(x,Q^2)/d\ln Q^2$. It is the main purpose of the study.

%However, the 
The $k_T$-factorization approach relates strongly to Regge-like behavior of 
parton distributions.
So, we restrict our investigations to 
%a Regge-likeform of 
SF
%structure functions 
and parton distributions at the following form (hereafter $a=q,g$):
%, one obtains (see \cite{KoMPL})
\begin{eqnarray} f_a(x,Q^2) \sim F_2(x,Q^2) \sim  x^{-\delta(Q^2)} 
\label{1.9} \end{eqnarray}

Note that really the slopes of the sea quark and gluon distributions:
$\delta_q$ and $\delta_g$, respectively, and the slope $\delta_{F2}$
of SF $F_2$ are little different.
The slopes have a familiar property $\delta_q <\delta_{F2} 
<\delta_g $ (see Refs. \cite{H1slo}-\cite{KoMPL} and references therein).
We will neglect, however, this difference and use
%Moreover, in %In 
in our investigations the experimental values of 
$\delta(Q^2) \equiv \delta_{F2}(Q^2)$ 
extracted by H1 
%and ZEUS 
Collaboration 
\footnote{ Now the preliminary ZEUS data for the slope $d\ln F_2/d\ln (1/x)$
are available as some points on Figs. 8 and 9 in Ref. \cite{slope}.
Moreover, the new preliminary H1 points have been presented on the 
Workshop DIS2002 (see \cite{DIS02}). Both the new points
%They 
are shown quite similar properties to compare with H1 data 
\cite{H1slo}. Unfortunately, tables of the ZEUS data and the new H1 data are
 unavailable yet
and, so, the points cannot be used here.} (see \cite{H1slo}
and references therein).
We 
%would like to 
note that the $Q^2$-dependence is in very good agreement
with perturbative QCD at $Q^2 \geq 2$ GeV$^2$ (see \cite{KoPaslo}). 
Moreover,
the values of the slope $\delta(Q^2)$ are 
in agreement with recent phenomenological studies (see, for example,
\cite{BFKLP}) incorporating the next-to-leading corrections \cite{FaLi}
(see also \cite{KoLi})
in the framework of BFKL approach.\\

 Thus, assuming the {\it Regge-like behavior} (\ref{1.9})
for the gluon distribution and
 $F_2(x,Q^2)$ at $x^{-\delta} \gg 1$
and using the NLO 
approximation for collinear coefficient functions and anomalous
dimensions of Wilson operators,
%of $r^{1+\delta}_{sp}$ and $r^{1+\delta}_{Lp}$
the following results for F$_L(x,Q^2)$ has been obtained in \cite{KoPaFL}:
\begin{eqnarray} \z  F_L(x, Q^2)  = - 2 
\frac{B^{g,1+\delta}_L \Bigl(1+ a_s(Q^2) \overline R^{g,1+\delta}_L \Bigr)}{ 
\gamma^{(0),1+\delta}_{qg} + 
\overline \gamma^{(1),1+\delta}_{qg} a_s(Q^2) } \xi ^{\delta}  
\biggl[
\frac{d F_2(x \xi, Q^2)}{d\ln Q^2} \nonumber \\ 
\z  ~~ ~~~~~~~~~~ ~+~ \frac{a_s(Q^2)}{2}  
\biggl( \frac{B^{q,1+\delta}_L}{B^{g,1+\delta}_L} 
\gamma^{(0),1+\delta}_{qg} - \gamma^{(0),1+\delta}_{qq} \biggr) 
%\xi ^{\delta} 
F_2(x \xi,Q^2) \biggr]
%&+&
+
O(a_s^2,x^{2-\delta},\alpha x^{1-\delta}), 
\label{9.1} 
\end{eqnarray}
where
$$\overline \gamma^{(1),\eta}_{qg} ~=~  \gamma^{(1),\eta}_{qg} + 
B_2^{q,\eta} \gamma^{(0),\eta}_{qg}
+
B_2^{g,\eta} \bigl(2\beta_0 + 
  \gamma^{(0),\eta}_{gg} - \gamma^{(0),\eta}_{qq} \bigr),~~
\overline R^{g,\eta}_L ~=~ R^{g,\eta}_L - B^{g,\eta}_2 
\frac{B^{q,\eta}_L}{B^{g,\eta}_L} $$

Here $\overline \gamma^{(1),\eta}_{qa} $ and $\overline R^{a,\eta}_L$
$(a=q,g)$ are the combinations \footnote{Because
  we consider here $F_2(x,Q^2)$ but not the singlet quark
  distribution in the corresponding DGLAP equations \cite{DGLAP}.} 
of the 'anomalous dimensions' 
%(AD) 
of Wilson operators 
$\gamma^{\eta}_{qa}= a_s \gamma^{(0),\eta}_{qa} + a_s^2 
\gamma^{(1),\eta}_{qa} + O(a_s^3)$
and 'Wilson coefficients'
$ a_s B_L^{a,\eta} 
\Bigl(1+ a_s R_L^{a,\eta} \Bigr)  + O(a_s^3)$
and $ a_s B_2^{a,\eta}  + O(a_s^2)$
with the 'moment' argument $\eta$  (i.e., the combinations of the functions
which can be obtained by analytical continuation of the 
corresponding anomalous dimensions and coefficient functions
%to variables extended
from 
%corresponding 
integer values $n$ of their argument  to non-integer ones $\eta$).

Note that, in principle, any term like $\sim 1/(n+m)$ $(m=0,1,2, ...)$
which 
comes to the corresponding combinations of the 
anomalous dimensions 
%(AD) 
and coefficient functions: $\overline \gamma^{(1),n}_{qa} $ and 
$\overline R^{a,n}_L$, 
should contribute to Eq.(\ref{9.1})
in the following form 
%\cite{method1} 
(after replacement of Mellin convolutions by usual products in the 
%corresponding 
DGLAP equations (see \cite{method1})):
\begin{eqnarray} 
\frac{1}{1+\delta + m} \, \left( 1 ~+~ 
\frac{\Gamma{(2+\delta + m)}\Gamma{(1+\nu)}}{\Gamma{(2+\delta + m+1+\nu)}}
\cdot x^{1+\delta + m} \right) \,, 
\label{d1.1} \end{eqnarray}
where the value of $\nu$ comes (see \cite{YF93,TMF,JeKoPa})
from asymptotics  of parton distributions 
$f_a(x)$ at $x \to 1$: $f_a \sim (1-x)^{\nu_a}$,
and 
\footnote{In our 
%small $x$ 
formula (\ref{9.1}) we have our interest
mostly to gluons, so we can apply $\nu =\nu_g \approx 4$ below.}
$\nu \approx 4$ from quark account rules \cite{schot}.
The additional term $\sim x^{1+\delta + m}$ in Eq. (\ref{d1.1}) is important 
only for $m=-1$ case (i.e. for the singular parts $\sim 1/(n-1)$ of the
corresponding anomalous dimensions and coefficient functions)
and quite small values of $\delta $ (i.e. for 
$x^{\delta} \sim Const$). 

Thus, excepting the case when $m=-1$ and $x^{\delta} \sim Const$, we
can replace 
%the term 
(\ref{d1.1}) by its first term
%the one 
$ 1/(1+\delta +m)$, i.e.
our variables $\overline \gamma^{(1),\eta}_{qg}$ and 
$\overline R^{g,\eta}_L$ are just 
%an analytical continuations of
the  combinations of corresponding anomalous dimensions and coefficient 
functions at $n=1+\delta$. 
When $m=-1$ and 
%the property 
$x^{\delta} \sim Const$ 
%is correct
at the considered small $x$ range, we should replace the term 
%$1/(n-1)=1/\delta$ in the anomalous dimensions and coefficient 
%functions at $n=1+\delta$
$1/(n-1)$,
if it was exist in the variables  $\overline \gamma^{(1),n}_{qa} $ and 
$\overline R^{a,n}_L$ $(a=q,g)$, 
%into Eq.(\ref{9.1}), 
by the following term:
\begin{eqnarray}
%\frac{1}{\delta} \stackrel{\delta \to 0}{\to} 
\frac{1}{\tilde \delta}
&=& \frac{1}{\delta}
\biggl[
1 ~-~ \frac{\Gamma(1- \delta)\Gamma(1+ \nu)}{\Gamma(1- \delta + \nu)} 
x^{\delta} \biggr]
\label{d1.2} \end{eqnarray}

%We would like to note 
Note also that the $1/\tilde \delta$
coincides approximately with $1/\delta$ when $\delta \neq 0$ and 
$x \to 0$. However, at $\delta \to 0$, the value of $1/\tilde \delta$ is not 
singular:
\begin{eqnarray}
\frac{1}{\tilde \delta} 
\to 
\ln \left(\frac{1}{x}\right) - \biggl[ \Psi(1+ \nu) - \Psi(1) \biggr] 
\label{10.7} \end{eqnarray}

So, the Eq. (\ref{9.1}) 
%and (\ref{9.2}) 
together with well-known 
expressions of 
%AD
anomalous dimensions $\gamma^{(0),n}_{ab} $ and 
$\gamma^{(1),n}_{ab}$ $(a,b=q,g)$ 
and coefficient functions 
$B^{a,n}_2 $ and 
$R^{a,n}_L$ $(a=q,g)$  (see
\cite{FlKoLa} and \cite{KaKo,KaKoFL}, respectively, and references therein) 
gives a possibility to extract SF $F_L$ at small $x$ values. The calculations
are based on precise experimental data of SF $F_2$ and its derivatives
$dF_2/d\ln(Q^2)$ and $\delta \equiv d\ln F_2/d\ln(1/x)$. 

For concrete $\delta$-values, the Eq. (\ref{9.1}) 
%and (\ref{9.2}) 
simplifies
essentially (see \cite{KoPaFL}). For example, for
%of $\delta = 0.5$ and 
$\delta = 0.3$ we obtain (for the number of active quarks $f=4$ and 
$\overline{MS}$ scheme):
\begin{eqnarray} 
\z  
F_L(x, Q^2)  = \frac{0.84}{1 + 59.3 a_s(Q^2) }  \biggl[
 \frac{d F_2(0.48 x, Q^2)}{d\ln Q^2} + 3.59 a_s(Q^2) 
F_2(0.48 x , Q^2) \biggr]
\nonumber \\ \z
~+~ 
O(a_s^2 \!\!, x^{2-\delta} \!\!, a_s x^{1-\delta}) 
\!         
\label{10.3}
\end{eqnarray}

At arbitrary $\delta$ values,
in real applications it is very useful to simplify 
Eq. (\ref{9.1}) 
%and (\ref{9.2}) 
as follows. We keep
the exact $\delta $-dependence only for the 
leading order 
%(LO) 
terms, which are very simple.
In the NLO corrections we extract the terms $\sim 1/\tilde \delta $, which
%are singular at $\delta \to 0 $
are changed strongly when $0 \leq \delta \leq 1$, and parameterize the rest
terms in the form:
$ a_i +b_i\delta  +c_i\delta ^2 $.
The coefficients
$a_i,~ b_i,~c_i$ are fixed from the agreement these parameterizations
%$ a_i +b_i\delta  +c_i\delta ^2 $ 
with the exact values of $\overline \gamma^{(1),\eta}_{qa} $ and 
$\overline R^{a,\eta}_L$
%for the terms
at $\delta=0,~0.3 $ and $0.5$.
These exact values can be found in Refs. \cite{KoPaG,KoPaFL,KoPaR}.

Then, the approximate representation of Eq. (\ref{9.1}) 
%and (\ref{9.2})
of  for arbitrary $\delta $ value has the form:
\begin{eqnarray}   
\z  F_L(x, Q^2)  = 
%\frac{4(1+ \delta)}{(4+ 3\delta + \delta^2)}
\frac{r(1+\delta){(\xi(\delta))}^{\delta}}{ (1 + 30 a_s(Q^2) [
1/\tilde \delta - \frac{116}{45} \rho_1(\delta)])  } \biggl[
\frac{d F_2(x\xi(\delta), Q^2)}{d\ln Q^2} 
+ \frac{8}{3} \rho_2(\delta) a_s(Q^2)  F_2(x\xi(\delta), Q^2)   \biggr] 
\nonumber \\ \z
~+~ 
O(a_s^2, a_s x, x^2), 
\label{10.51} 
\end{eqnarray} 
where
\begin{eqnarray}
%\frac{1}{\delta} \stackrel{\delta \to 0}{\to} 
r(\delta) &=& \frac{4\delta}{2+\delta+\delta^2},~~~
\xi(\delta) ~=~ \frac{r(\delta)}{r(1+\delta)},\nonumber \\
\rho_1(\delta) &=& 1+ \delta + \delta^2/4,~~~
\rho_2(\delta) ~=~ 1-2.39 \delta + 2.69 \delta^2, 
\label{10.6} \end{eqnarray}

%%%%%%%%%%%%%%%%%%%%%%%%  4 4 4 

\section{Comparison with $F_L$ experimental data } \indent

%\vskip 0.5cm
With the help of the results obtained in the previous Section
we have analyzed experimental data for SF $F_L$ 
\footnote{Sometimes there are experimental data for the ration 
$R=\sigma_L/\sigma_T$, which can be recalculated for the
%transformed to the data of 
SF $F_L$ because $F_L=F_2 R/(1+R)$.}
from H1 \cite{H1FL97} and \cite{Gogi},
NMC \cite{NMC}, 
%SLAC \cite{SLAC}, 
CCFR \cite{CCFR} and \cite{CCFRr},
BCDMS \cite{BCDMS} collaborations
\footnote{ We do not use experimental data of $R$ from SLAC 
\cite{SLAC,SLAC1}, EM \cite{EMC} and
CDHSW \cite{CDHSW} Collaborations because they are obtained at quite large
$x$ values.}.
Note, that we do not correct CCFR data \cite{CCFR} and \cite{CCFRr} which
has been obtained in $\nu N$ processes because the terms 
$\sim x \cdot m_c^2/Q^2$,
which are different in $\mu N$ and $\nu N$ processes,
are not so strong at low $x$. 

We calculate the SF $F_L$ as the sum of two types of contributions:
charm quark one $F^c_L$ and light quark one $F^{l}_L$:
\bea 
F_L  ~=~ F^{l}_L + F^{c}_L
\label{nu1}
\eea
We use the expression  (\ref{d1}) 
for the calculation of the both SF $F^{l}_L$ and $F^{c}_L$ 
in the  following form (here $a_c=m^2_c/Q^2$):
\bea
F^{l}_L(x, Q^2) &=& \sum_{f=uds} e^2_f \biggl[
\int_{x}^{1} \frac{dz}{z} \hat C_{L,BFKL}^g(\frac{x}{z}, Q^2, 0) \, 
zf_g(z, Q_0^2) +  
\nonumber \\
&+& \sum_{i=1}^2
\int_{z_{min}^{(i)}}^{z_{max}^{(i)}} \frac{dz}{z} \int_
{k_{{\bot}min}^{2(i)}}^{k_{{\bot}max}^{2(i)}} dk_{\bot}^2 \,
\hat C_{L,BFKL}^g(\frac{x}{z}, Q^2, k_{\bot}^2) \, 
\Phi (z, k_{\bot}^2, Q_0^2)\biggl], 
\label{zo4}\\
F^{c}_L(x, Q^2) &=& e^2_c \biggl[
\int_{x(1+4a_c)}^{1} \frac{dz}{z}
\hat C_{L,BFKL}^g(\frac{x}{z}, m^2_c, Q^2, 0) \, zf_g(z, Q_0^2) +  
\nonumber \\ &+& \sum_{i=1}^2
\int_{z_{min}^{(i)}}^{z_{max}^{(i)}} \frac{dz}{z} \int_
{k_{{\bot}min}^{2(i)}}^{k_{{\bot}max}^{2(i)}} dk_{\bot}^2 \,
\hat C_{L,BFKL}^g(\frac{x}{z},m^2_c, Q^2, k_{\bot}^2) \, 
\Phi (z, k_{\bot}^2, Q_0^2)\biggr], 
\label{zo4.1}
\end{eqnarray}
%Here $\Phi (y, k_{\bot}^2, Q_0^2) = \frac{1}{k_{\bot}^2} \varphi_{g}
%(y, k_{\bot}^2, Q_0^2)$ and the changes $x_{B} \to x, x \to y$ were
%done in comparison with (\ref{d1}). 
where
$\hat C_{L,BFKL}^g(x_B, Q^2, m^2_c, k_{\bot}^2)$ are given by 
Eqs. (\ref{4}) and (\ref{4dd}).

The  integration limits in the expression (\ref{zo4.1}) have the following
values:
\begin{eqnarray}
z^{(1)}_{min} &=& x(1 + 4a_c + \frac{Q_0^2}{Q^2}), \,\,  z^{(1)}_{max} ~=~
2x(1 + 2a_c);  \nonumber \\
k_{{\bot}min}^{2(1)} &=& Q_0^2, \,\, k_{{\bot}max}^{2(1)} ~=~
(\frac{z}{x} - (1 +4a_c))Q^2; \nonumber \\
z^{(2)}_{min} &=& 2x(1 + 2a_c), \,\, z^{(2)}_{max} ~=~ 1; \nonumber \\
k_{{\bot}min}^{2(2)} &=& Q_0^2, \,\, k_{{\bot}max}^{2(2)} ~=~ Q^2; 
\label{ranges}
\end{eqnarray}

The ranges of integration correspond to positive values of
square roots in expressions  (\ref{5}), (\ref{6}), 
(\ref{5dd}) and (\ref{6dd})
and also should obey to kinematical restriction 
$(z \leq (1+4a_c+b)^{-1}$) following from condition $\beta^2 \geq 0$ (see Eqs.
(\ref{5})-(\ref{6.1d})).
%  x/y) \leq z_0$ with $z_0$ from Eq.(\ref{1A}). 
%We would like to note that 
In Eq.(\ref{zo4}) the ranges 
(\ref{ranges}) are used at $a_c=0$.\\

In Fig. 2 we show  the SF $F_L$ as a function $x$ for different values
of $Q^2$  in comparison with H1 experimental data sets: old one of
\cite{H1FL97} 
(black triangles), last year one of \cite{H1FL} (black squares) and 
new preliminary one of \cite{Gogi} (black circles) and also with NMC 
\cite{NMC} (white triangles), CCFR \cite{CCFR} (white circles) and BCDMS
\cite{BCDMS} data (white squares).
For comparison with these data
we present the results of the calculation with three 
different 
parameterizations for the unintegrated gluon distribution 
$\Phi (x, k^2_{\bot}, Q_0^2)$ 
%in the forms given by Eq. (\ref{zo1})\, and Eq. (\ref{zo1dd}) 
at $Q_0^2 = 4$ GeV$^2$. 
%{\bf luchshe by dobavit' nazvaniya unint. gl. raspred. i ssylki}.
%Two 
All of them: 
%Ryskin-Shabelsky (RS) 
Kwiecinski-Martin-Stasto (KMS) one  
%from 
\cite{RySha} 
and Blumlein (JB) one \cite{Blumlein} and Golec-Biernat and Wusthoff (GBW), 
have been used already in our 
previous work \cite{KLPZ} and reviewed there.
% (see also \cite{LiZo}).
%The third unintegrated gluon function used here is one proposed by
%Golec-Biernat and Wusthoff (GBW) which takes into account
%saturation effects and has been applied earlier in analysis of the 
%inclusive and diffractive $ep$-scattering data \cite{GoBiWu}. 
%%It hast the following form:\\

There are several other popular parameterizations (see, for example,
Kimber-Martin-Ryskin (KMR)  \cite{KiMaRy} and Jung-Salam (JS)  
\cite{JuSa}),
which are not used in our study mostly because of technical difficulties.
%\footnote{Note that the RS parameterization \cite{RS} is quite old. 
%We use it together with the JB set \cite{BL} (when the value 
%$\mu^2 = Q_0^2 = 1$ GeV$^2$) only one time (see Fig. 6) to prove
%the coincidence between our off mass shell matrix elements and those from 
%Ref.~\cite{BMSS}.}. 
Note that all above parameterizations give quite similar results  excepting, 
perhaps, the contributions from the small $k_{\bot}^2$-range: 
$k_{\bot}^2 \leq 1$ GeV$^2$ 
(see Ref. \cite{Andersson} and references therein). Because we use 
$Q_{0}^2 =4$ GeV$^2$
in the study of SF $F_{L}$, our results depend very slightly  on 
the the small $k_{\bot}^2$-range of the parameterizations.
In the case JB, GBW and KMS sets this observation is supported
below by our results and we expect that the application of KMR and JS sets
should not strongly change our results.

 The differences observed between the curves 2, 3 and 4 are 
 due to the different behavior of the unintegrated gluon distribution
as function $x$ and $k_{\bot}$.
 We see that 
%at large $Q^2$ ($Q^2 \geq 7$ GeV$^2$)
the SF $F_L$
obtained in the $k_T$-factorization approach with KMS and JB parameterizations
is close each other 
\footnote{Note that very similar results
have been obtained also for Ryskin-Shabelsky parameterization \cite{RySha} 
(see \cite{KoLiZo01}).}
and higher than the SF obtained
in the pure perturbative QCD
%standard parton model 
with the GRV 
%and MT ~\cite{MT} 
gluon density at the leading order approximation.
Otherwise, the $k_T$-factorization approach with GBW parameterization
is very close to pure QCD predictions
\footnote{This 
fact is evident also from quite large value of $Q^2_0=4$ 
GeV$^2$ chosen here.}: 
it should be so because 
GBW model has deviations from perturbative QCD only at quite low $Q^2$
values.
Thus, the predictions of perturbative QCD 
%(in GRV approach)
and ones based on $k_T$ factorization approach are in
%quite similar and show a good 
agreement each other and with all data 
within modern experimental uncertainties. So, a possible high values
of high-twist corrections to SF $F_L$ predicted in \cite{BaGoPe}
can be important only at 
%quite
low $Q^2$ values: $Q^2 \leq Q^2_0=4$ GeV$^2$.\\

Fig. 3 is similar to Fig. 2 with one excepting: 
%the first H1 data of $F_L$ \cite{H1FL97} are replaced by 
we add 'experimental data' obtained using the relation 
between SF $F_L(x,Q^2)$, $F_2(x,Q^2)$
and $dF_2(x,Q^2)/d\, {\ln Q^2}$ (see Section 3) as black stars. Because 
the corresponding data for 
SF $F_2(x,Q^2)$ and $dF_2(x,Q^2)/d\, {\ln Q^2}$ essentially more precise
(see \cite{H1FL})
to compare with the preliminary data \cite{Gogi} for $F_L$, 
%the uncertainties of 
the 'experimental data' have strongly suppressed uncertainties.
%are essentially less.
As it is shown on Fig. 3 there are very good agreement between the new
preliminary data \cite{Gogi}, the 'experimental data' and predictions
of perturbative QCD and $k_T$-factorization approach.\\

To estimate of the value of charm mass effect, we recalculate SF $F_L^c$
%and $F_L^c$ 
also in massless approximation similar to (\ref{zo4}). 
 In Fig. 4 we show importance of exact $m_c$-dependence in hard SF of 
%charm part of 
$F_L^c$ to compare with its massless approximation, where we should have
%the ration
$F_L^c(m_c=0)/F_L(m_c=0) =2/5$. As it is possible to see on Fig. 4, 
the ratio $F_L^c/F_L $ goes to massless limit $2/5$ only at
asymptotically large $Q^2$ values.

\section{Conclusions} \indent

We have applied in the framework of $k_T$-factorization approach the
%The application of our 
results of the calculation of the perturbative
parts for the structure functions $F_L$ and $F_L^c$ 
for a gluon target 
having nonzero momentum square, in the process 
of photon-gluon fusion
to the analysis of
present data for the structure function $F_L$
\footnote{In Ref \cite{KLPZ} we have also obtained
quite large contribution of SF $F_L^c$ at low $x$  and
high $Q^2$ ($Q^2 \geq 30$ GeV$^2$).}.
%have been given.
The analysis has been performed with 
several parameterizations of unintegrated gluon
distributions, for comparison. We have found good agreement
%, perhaps, except at low $Q^2$ ($Q^2 \leq 7$ GeV$^2$), 
between all existing experimental data, 
%the preliminary data \cite{Gogi}, 
the predictions for 
$F_L$ obtained from the relation between SF $F_L(x,Q^2)$, $F_2(x,Q^2)$
and $dF_2(x,Q^2)/d\ln Q^2$
and the results 
obtained in the framework of 
perturbative QCD 
%(based on GRV approach)
and ones based on $k_T$-factorization approach
 with the three different 
%RS(?) and BFKL(?) 
parameterizations of unintegrated gluon
distributions.

We note that 
it could be also very useful to evaluate the SF $F_2$ itself
\footnote{A study of the SF $F_2$ in the framework of $k_T$-factorization
has been already done in Ref. \cite{Jung}.} and 
the derivatives of $F_2$ respect to the 
logarithms of $1/x$ and $Q^2$ with our expressions using the unintegtated
gluons.
%by these methodics. 
We are considering to present this work and also the predictions for the 
ratio $R=\sigma_L/\sigma_T$ in a forthcoming article.

The consideration of the SF $F_2$ in the framework of the leading-twist
approximation of
perturbative QCD (i.e. for ``pure'' perturbative QCD)
leads to
very good agreement (see Ref. 
\cite{Q2evo} and references therein) with the HERA
data at low $x$ and $Q^2 \geq 1.5$ GeV$^2$. The agreement improves
%become to be better
at lower $Q^2$ when higher twist terms are 
taken into account in Ref. \cite{HT,KriKo}. As
it has been studied in Refs. \cite{Q2evo,HT}, the SF $F_2$  at low 
$Q^2$ 
%values 
is sensentive to the
small-$x$ behavior of quark distributions.
Thus, our future
analysis of $F_2$ 
in broad $Q^2$ range 
in the framework of $k_T$-factorization approach
should require the incorporation of parameterizations of unintegrated
quark densities, introduced recently (see Ref. \cite{KiMaRy,Andersson} and 
references therein).\\

%
%
%\vspace{1cm}
\hspace{1cm} \Large{} {\bf Acknowledgements}    \vspace{0.5cm}

\normalsize{}

We are grateful to 
Professor Catani for useful discussions and comments.

The study is supported in part by the RFBR grant 02-02-17513.
One of the authors (A.V.K.) is supported in part
by Heisenberg-Landau program
%Alexander von Humboldt fellowship 
and INTAS  grant N366.
%A.V.L. is supported in part by INTAS  grant (?).
N.P.Z. also
   acknowledge the support of Royal Swedish Academy of Sciences.

%\newpage

%\newpage

%\vspace{0.5cm}

%\hspace{1cm} {\Large{\bf Figure captions}}    \vspace{0.5cm}

%\end{document}
\newpage

\vspace{0.5cm}

\hspace{1cm} {\Large{\bf Figure captions}}\\
\vspace{0.5cm}

\noindent {\bf Fig.~1~}
The diagrams contributing to $T_{\mu\nu}$ for a gluon target.
They should be multiplied by a factor of 2 because of the opposite direction 
of the fermion loop. The diagram (a) 
should be also doubled because of crossing symmetry.\\

\noindent {\bf Fig.~2~}
The structure function $F_L(x,Q^2)$ as a function of $x$ for
different values of $Q^2$ compared to experimental data. The H1 data:
the first 1997 ones ~\cite{H1FL97},
new 2001 ones ~\cite{H1FL} and preliminary ones ~\cite{Gogi}
are shown as black triangles, circles and squares, respectively.
The data of NM \cite{NMC}, CCFR \cite{CCFRr} and BCDMS \cite{BCDMS} 
Collaborations
are shown as white triangles, circles and squares, respectively.
Curves 1, 2, 3 and 4 correspond to SF obtained
in the perturbative QCD with the GRV ~\cite{GRV}
gluon density at the leading order approximation and to SF
obtained in the $k_T$ factorization approach with 
%RS~\cite{RySha}, 
JB (at $Q_0^2 = 4$ GeV$^2$)~\cite{Blumlein},
Kwiecinski-Martin-Stasto (KMS)
 and GBW \cite{GoBiWu} 
parametrizations of unintegrated gluon distribution.\\

\noindent {\bf Fig.~3~}
The structure function $F_L(x,Q^2)$ as a function of $x$ for
different values of $Q^2$. To compare with Fig. 2 'experimental data'
(see \cite{KoFL}-\cite{KoPaR} and Section 3) are added as black stars.
The 'experimental data' values depend mostly on the derivative 
$dF_2(x,Q^2)/d\, {\ln Q^2}$, which data are know at little different $Q^2$
values (see \cite{H1FL}). So, the 'experimental data' obtained at 
%$Q^2$ values (in GeV$^2$): 
12, 15, 20, 25 and 35 GeV$^2$ are presented here at 
13.4, 15.3, 22.4, 29.6 and 39.7 GeV$^2$, respectively.\\

\noindent {\bf Fig.~4~}
The ratio $F_L^c(x,Q^2)/F_L(x,Q^2)$ as a function of $x$ for
different values of $Q^2$.\\

\newpage

\begin{figure}
\begin{center}
\epsfig{figure=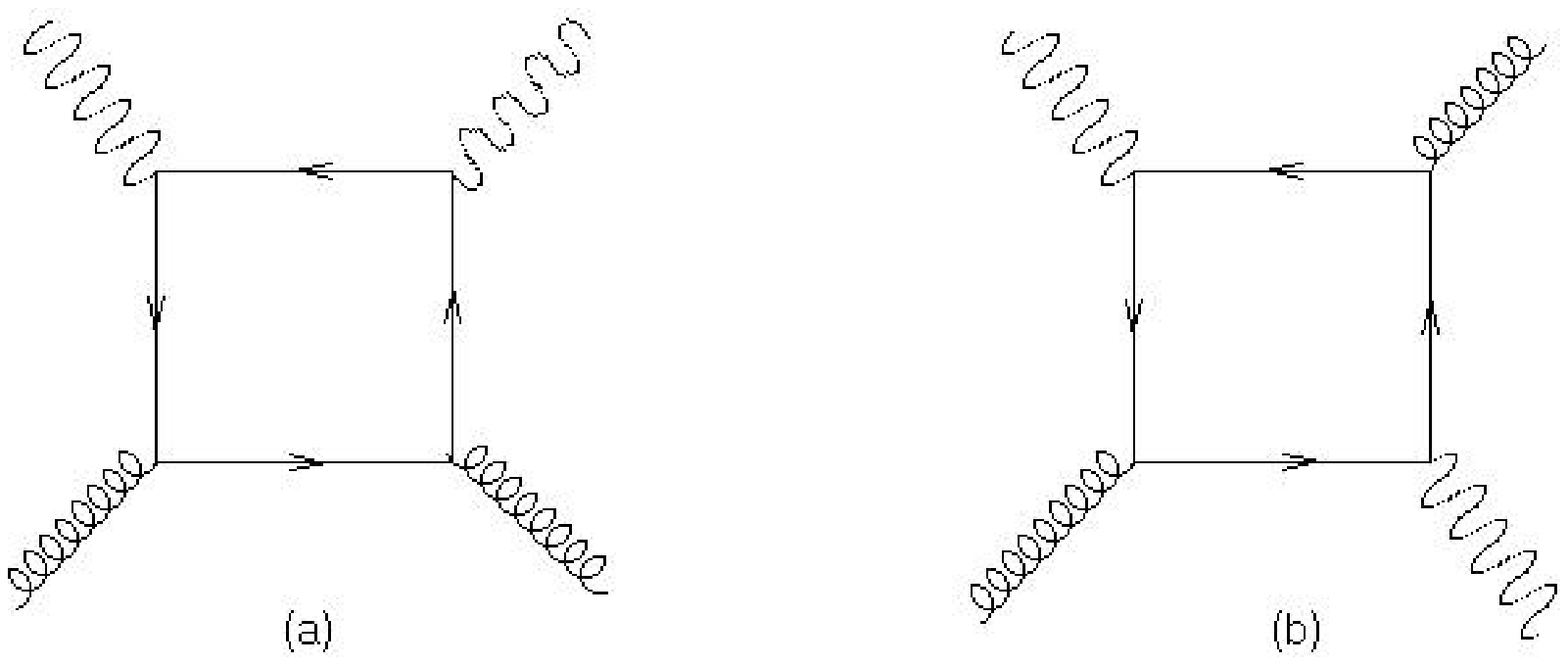,width=16cm,height=12cm}
\end{center}
\caption{}
%The box diagramms of photon gluon fusion.}
\label{fig1}
\end{figure}

\begin{figure}
\begin{center}
\epsfig{figure=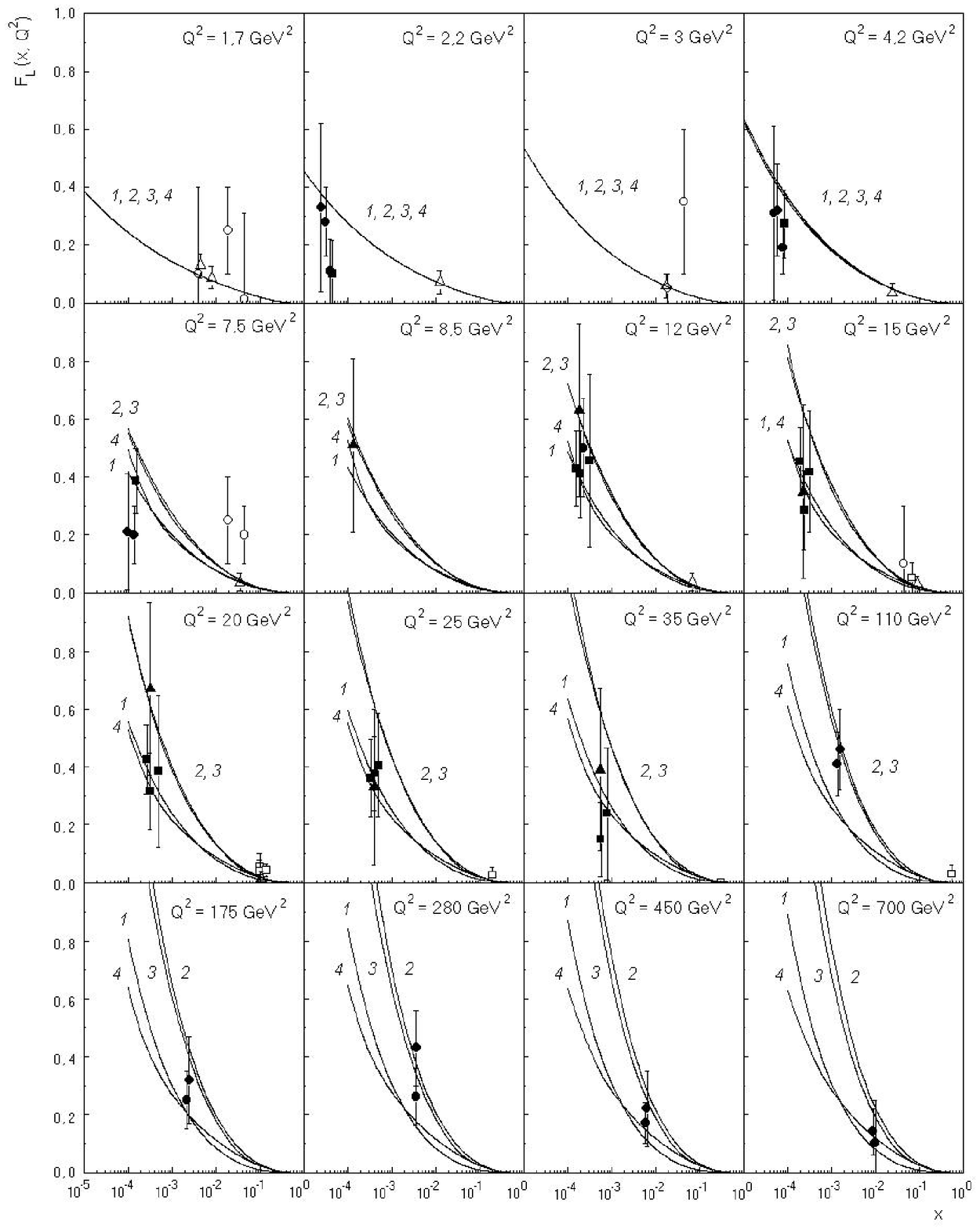,width=16.5cm,height=23.5cm}
\end{center}
\caption{}
%\vspace{-1cm} 
\label{fig2}
\end{figure}

\begin{figure}
\begin{center}
\epsfig{figure=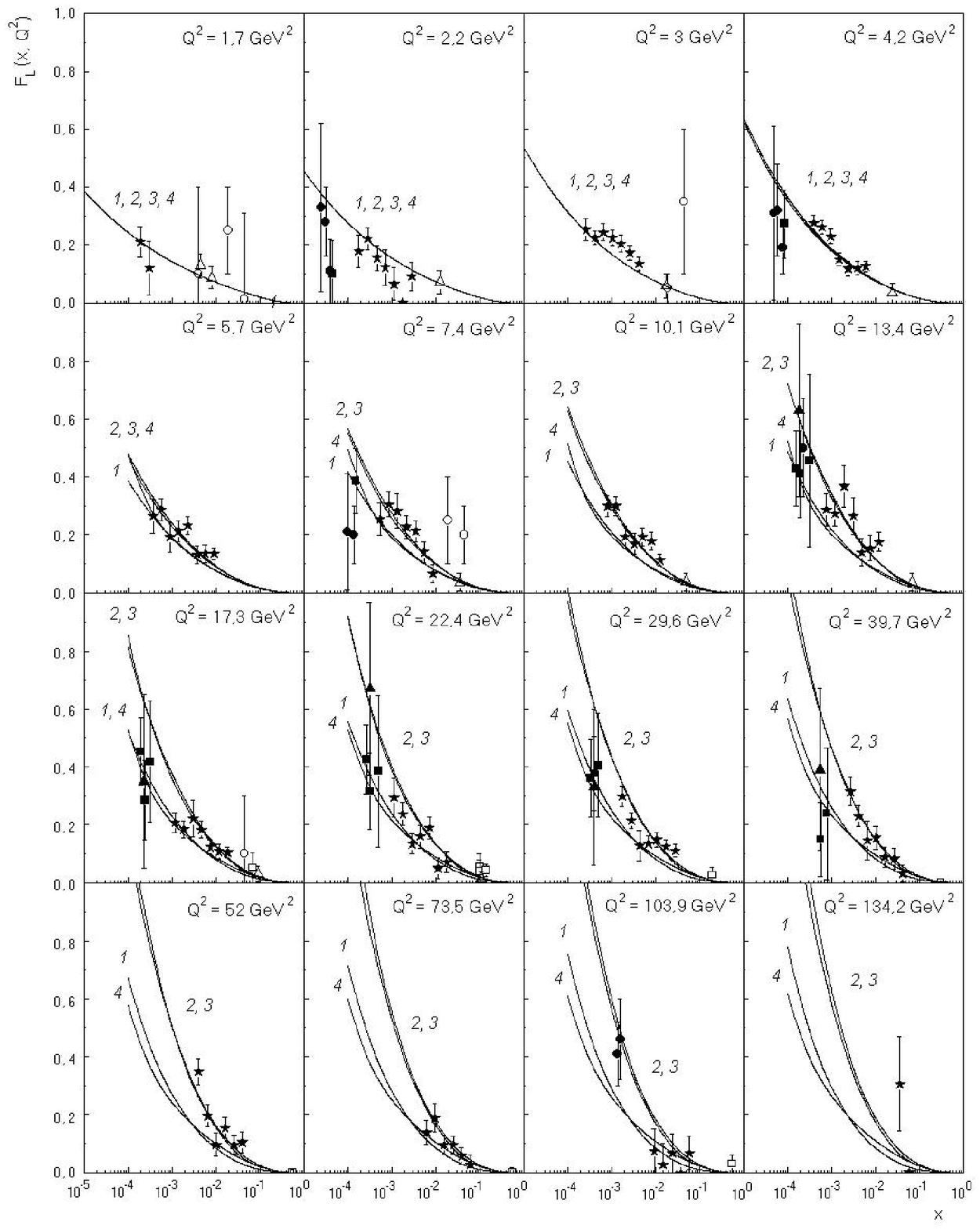,width=16.5cm,height=23.5cm}
\end{center}
\caption{}
\label{fig3}
\end{figure}

\begin{figure}
\begin{center}
\epsfig{figure=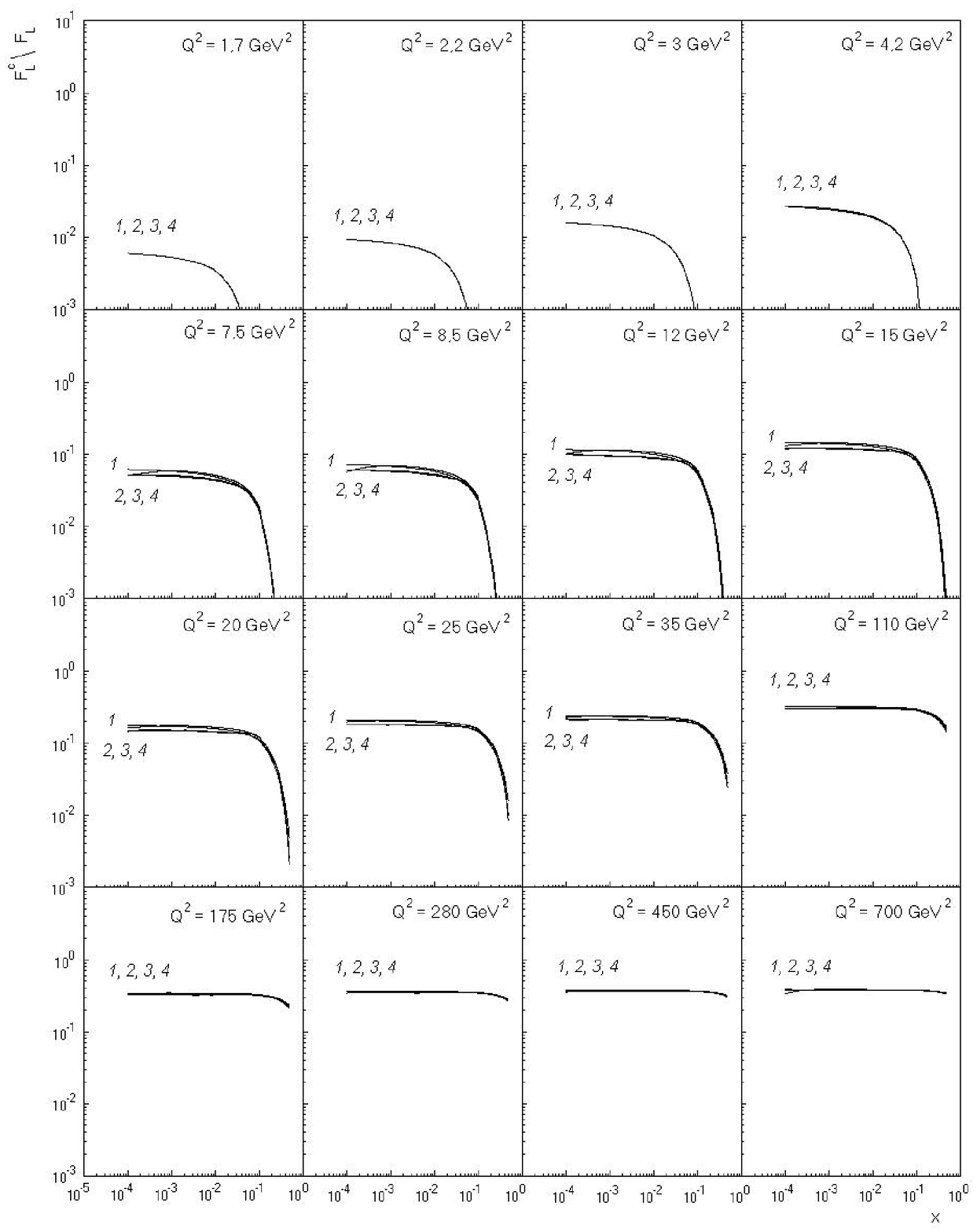,width=16.5cm,height=23.5cm}
\end{center}
\caption{}
\label{fig4}
\end{figure}
\end{document}